\documentclass[prd, amsfonts, twocolumn, nofootinbib, showpacs]{revtex4-2}
\usepackage{graphicx, epsfig}
\usepackage{color}
\usepackage{amsmath}
\usepackage{amssymb}
\usepackage{physics}
\newcommand{\be}{\begin{equation}}
\newcommand{\ee}{\end{equation}}
\newcommand{\bea}{\begin{eqnarray}}
\newcommand{\eea}{\end{eqnarray}}

\newcommand{\gapp}{\mathrel{\raise.3ex\hbox{$>$}\mkern-14mu
\lower0.6ex\hbox{$\sim$}}}
\newcommand{\lapp}{\mathrel{\raise.3ex\hbox{$<$}\mkern-14mu
\lower0.6ex\hbox{$\sim$}}}
\def\bbox{{\,\lower0.9pt\vbox{\hrule \hbox{\vrule height 0.2 cm
\hskip 0.2 cm \vrule  height 0.2 cm}\hrule}\,}}

\begin{document}
\title{Kerr black holes as circular polarizers }
\author{De-Chang Dai$^{1,2}$ }
\affiliation{ $^1$ Department of Physics, national Dong Hwa University, Hualien, Taiwan, Republic of China}
\affiliation{ $^2$ CERCA, Department of Physics, Case Western Reserve University, Cleveland OH 44106-7079}

\begin{abstract}
\widetext
We study the retrograde second caustics of extremal Kerr black holes, where the intensity of the light beam is infinitely magnified. We find that the caustics of different polarized beams are split by as much as $10^{-3}$rad by an external black hole for a suitable range of parameters. A lensing black hole at several lys away separates the polarized beams about $10^{12}$m apart.  This splitting is larger than the radius of the Earth. Therefore, an observer on Earth would see different circularly polarized light according to their location. The polarization will change while the detector is wandering around. Thus, the polarization of light beams can be an important quantity in retrolensing observations.  
\end{abstract}


\maketitle

\section{introduction}
One of the fascinating cosmic phenomena predicted by General Relativity is the gravitational lensing, under which light is deflected and creates multiple or twisted images while passing by a massive object\cite{2008AHES...62....1S}. The basic theory of gravitational lensing was developed in the last century \cite{Liebes:1964zz,Refsdal:1964yk,1975ApJ...195...13B,Blandford:1991xc,1992grle.book.....S,Refsdal:1993kf,Narayan:1996ba,Wambsganss:1998gg}. Nowaday, gravitational lensing has become an important tool to study dark matter, extrasolar planets and cosmological parameters\cite{2002P&SS...50..299D,Sutherland:1998bm}. These studies mainly focus on weak gravitational fields. The deflected angle is approximated by $\alpha\approx 4M/D$ where $D$ and $M$ are the impact parameter and mass of the lens object respectively. Meanwhile, it has been noticed that a photon can wind around a black hole several times without falling into the black hole horizon\cite{1959RSPSA.249..180D,1965AJ.....70..517A,Luminet:1979nyg,Ohanian:1987xga,Nemiroff:1993he}. This is a possible method to study strong gravitational fields near a black hole. 

In recent years, Holz and Wheeler revisited retrolensing and proposed that sunlight will be redirected back to Earth if a few solar mass black hole passes within $1$pc of the solar system\cite{Holz:2002uf}. An observer will see a star light up and dim gradually. It was also proposed that the star S2 orbiting Sgr A* black hole can be a suitable candidate for retrolensing object, and its nearby star S2 can be the light source\cite{DePaolisi2003,Eiroa:2003jf,DePaolis:2004xe}. The magnification and caustics have been studied intensively to search for the best observation requirement\cite{Bozza:2008mi,Bozza:2006nm,Bozza:2005kq,Bozza:2002af,Gyulchev:2006zg,Bozza:2004kq,Sereno:2007gd} and extended to modified gravities\cite{Zakharov:2004sg,Tsukamoto:2016oca}. Retrolensing can be a practical tool to study strong gravity since the distinguishable light paths in different theories.

\begin{figure}[h]
\includegraphics[width=5cm]{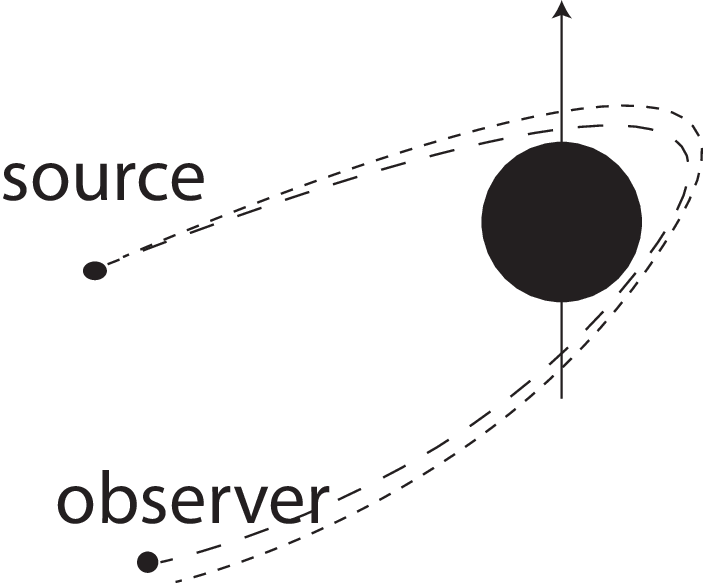}
\caption{A source emits non-polarized light. The unpolarized light is split into two polarized lights while passing a rotating black hole. Observers will see distinguishable polarization images based on their positions.     }
\label{light-path}
\end{figure}

In general, a test particle moves along its geodesic. This statement, however, must be corrected if spin is considered\cite{Mathisson,Dixon,Papapetrou}. The spin of an electromagnetic wave is $\pm \hbar$. The light beams of different polarizations are deflected by a black hole to directions separated by an angle 
\begin{equation}
\delta\sim \frac{\lambda J}{ D^3}
\end{equation}
in the weak gravitational fields approximation\cite{Mashhoon:1993,Mashhoon:1974,Mashhoon:1975}. Here, $\lambda$ and $J$ are the wavelength of light and the angular momentum of a black hole respectively. An unpolarized light beam from the same source is split into two polarized beams after passing a black hole (Fig.\ref{light-path}). Observers at different locations will see different combinations of polarized light rays. In other words, natural light becomes elliptically polarized. This is called spinoptics\cite{Frolov:2011mh,Frolov:2012zn,Frolov:2024ebe,Frolov:2024olb} and it has been extended to gravitational waves\cite{Kubota:2023dlz,Frolov:2024qow,Kubota:2024zkv}. 

There is no doubt that black holes deflect differently polarized light rays differently. The problem is how and where we can see the effect. Recent studies have found that isolated black holes do exist\cite{OGLE:2022gdj,2023ApJ...955..116L,2022ApJ...933L..23L}. These isolated black holes are good lensing objects for retrolensing. The other candidate is a primordial black hole, which may serve as a microlensing object. Another problem is how to overcome the brightness issue. The intensity of the light ray is infinitely magnified at caustics. Light near caustics may be a good example where to observe polarized beams splitting effect. Therefore, we study the caustics for different polarized lights. We find the splitting angle can be as big as $10^{-3}$rad for an extremal black hole. A lensing black hole at several lys away separates the polarized beams about $10^{12}$m apart, which is larger than the radius of the Earth. Thus the splitting should be observable for an observer on Earth. In particular, an observer at the right circular polarized caustic points sees strong right circular light and weak left circular light and vice versa. An observer may see different polarized beams while he moves through space. Since the isolated black hole distribution is not well studied, we cannot estimate the likelihood of finding such events. But isolated black holes do exist and have been found. It should be worth giving it a try. 

In the following, we first review the spin optics based on Frolov and Shoom's paper\cite{Frolov:2012zn}. The caustics and magnification are calculated numerically. Only the second retrograde caustics are presented since most of the prograde beam goes into mathematical singularities. Then we discuss the retrolensing based on the numerical result.

\section{Spin optics in the Kerr spacetime}
Spin optics is based on Maxwell's equations in curved spacetime. No modified theories have been involved. 
To study the path of the polarized light around a rotating black hole, we follow the numerical method given by Frolov and Shoom\cite{Frolov:2012zn}. We list only the required equations. For a deep study on the equations, please consult Frolov and Shoom's paper\cite{Frolov:2012zn} 

The rotating black hole spacetime is expressed in the Boyer-Lindquist coordinates as

\begin{equation}
ds^2=h\Big(-(dt-g_idx^i)^2+dl^2\Big) 
\end{equation}
, where 
\begin{eqnarray}
&&h=1-\frac{2Mr}{\Sigma} \text{, } g_i=-\frac{2aMr}{\Sigma h}\sin^2\theta \delta^\phi_i\\
&&\Sigma=r^2+a^2\cos^2\theta \text{, }\Delta =r^2-2Mr+a^2\\
&&dl^2=\frac{\Sigma}{\Delta h}dr^2+\frac{\Sigma}{h}d\theta^2+\frac{\Delta\sin^2\theta}{h^2}d\phi^2
\end{eqnarray}
where $M$ is the black hole mass and $a=J/M$ is the black hole spin parameter. $J$ is the angular momentum of the black hole.  
A new coordinate system $(\tau,x,y,\phi)$ is adopted to simplify the equations.
\begin{eqnarray}
&&\tau=\frac{t}{M}\text{, }x=\frac{M}{r}\text{, }y=\cos\theta\\
&&w_\pm=\tau\pm \int \frac{\eta}{x^2\delta}dx 
\end{eqnarray}
here 
\begin{eqnarray}
&&\eta=(1+2\alpha^2x^3+\alpha^2x^2)^{1/2}\\
&&\delta=1-2x+\alpha^2x^2\\
&&\alpha=\frac{a}{M} 
\end{eqnarray}

Then the Kerr metric is written as 

\begin{equation}
ds^2=M^2h \Big(-(d\tau^2-\gamma_i dz^i)^2 +\gamma_{xx}dx^2++\gamma_{yy}dy^2+\gamma_{\phi\phi}d\phi^2\Big)
\end{equation}
here, $z^i=(x,y,\phi)$, and the elements of the metric are 
\begin{eqnarray}
&&h=1-\frac{2x}{\sigma} \text{, } \gamma_i =-\frac{2\alpha x (1-y^2)}{(\sigma-2x)}\delta^\phi_i\\
&&\sigma=1+\alpha^2x^2y^2\\
&&\gamma_{xx}=\frac{\sigma^2}{x^4\delta (\sigma-2x)} \text{, }\gamma_{yy}=\frac{\sigma^2}{x^2(\sigma-2x)(1-y^2)}\\ 
&&\gamma_{\phi\phi}=\frac{\delta\sigma^2 (1-y^2)}{x^2(\sigma-2x)^2} \text{, }\gamma=\frac{\sigma^6}{x^8(\sigma-2x)^4}
\end{eqnarray}
The equations of motion of a polarized light are
\begin{eqnarray}
&&\frac{dx}{dw_{\pm}}=\frac{X}{A} \text{, } \frac{dy}{dw_{\pm}}=\frac{Y}{A},\\ 
&&\frac{dX}{dw_{\pm}}=-\frac{1}{2A\gamma_{xx}}(\gamma_{xx,x}X^2+2\gamma_{xx,y}XY-\gamma_{yy,x}Y^2)\nonumber\\
&&+\frac{\gamma_{\phi\phi ,x}}{2A\gamma_{xx}\gamma_{\phi\phi}^2}(p-\gamma_\phi)^2-\frac{(p-\gamma_\phi)}{A\sqrt{\gamma}}(\text{curl} \gamma)_y\nonumber\\
&&+\epsilon \frac{\gamma_{yy} (\text{curlcurl})_\phi}{A\sqrt{\gamma}}Y,\nonumber\\
&&\frac{dY}{dw_{\pm}}=-\frac{1}{2A\gamma_{yy}}(\gamma_{yy,y}Y^2+2\gamma_{yy,x}XY-\gamma_{xx,y}X^2)\nonumber\\
&&+\frac{\gamma_{\phi\phi ,y}}{2A\gamma_{yy}\gamma_{\phi\phi}^2}(p-\gamma_\phi)^2+\frac{(p-\gamma_\phi)}{A\sqrt{\gamma}}(\text{curl} \gamma)_x\nonumber\\
&&-\epsilon \frac{\gamma_{xx} (\text{curlcurl})_\phi}{A\sqrt{\gamma}}X,\nonumber\\
&&\frac{d\phi}{dw_{\pm}}=\frac{p-\gamma_\phi}{A\gamma_{\phi\phi}}
\end{eqnarray}
, where 
\begin{equation}
A=1+\frac{\gamma_\phi (p-\gamma_\phi)}{\gamma_{\phi\phi}} -\frac{\eta}{x^2\delta} |X|.
\end{equation}
$\epsilon=\pm\frac{1}{2\omega M}$. $\omega$ is the angular frequency of a photon. The $\pm$ sign defines the helicities of the photon; right circular light is $+$ and left circular light is $-$.
   
A null ray obeys, 
\begin{equation}
\gamma_{xx}X^2+\gamma_{yy}Y^2 +\frac{(p-\gamma_\phi)^2}{\gamma_{\phi\phi}}=1
\end{equation}
This conservation equation is used to control the accuracy of the numerical calculation. In our study, we keep the deviation smaller than $10^{-20}$. The above equation of motion is inconvenient when $x$ is close to $0$. The coordinates may be written as functions of variable $\phi$. The equation for $x(\phi)$ is
\begin{equation}
\frac{dx^2}{d\phi^2}=\tilde{A}\Big(\frac{dx}{d\phi}\Big)^2+\tilde{B}\frac{dx}{d\phi}\frac{dy}{d\phi}+\tilde{C}\Big(\frac{dy}{d\phi}\Big)^2+\tilde{D}\frac{dy}{d\phi}+\tilde{E}
\end{equation}

here 
\begin{eqnarray}
&&\tilde{A}=\frac{\gamma_{\phi\phi,x}}{\gamma_{\phi\phi}}-\frac{\gamma_{xx,x}}{2\gamma_{xx}}-\frac{\gamma_{xx}\gamma_{\phi\phi}}{(p-\gamma_{\phi})\sqrt{\gamma}}(\text{curl}\gamma)_y\nonumber\\
&&\tilde{B}=\frac{\gamma_{\phi\phi,y}}{\gamma_{\phi\phi}}-\frac{\gamma_{xx,y}}{2\gamma_{xx}}+\frac{\gamma_{yy}\gamma_{\phi\phi}}{(p-\gamma_{\phi})\sqrt{\gamma}}(\text{curl}\gamma)_x\nonumber\\
&&\tilde{C}=\frac{2\gamma_{yy,x}}{\gamma_{xx}}\nonumber\\
&&\tilde{D}=\epsilon\frac{\gamma_{yy}\gamma_{\phi\phi}}{(p-\gamma_{\phi})\sqrt{\gamma}}(\text{curlcurl}\gamma)_\phi\nonumber\\
&&\tilde{E}=\frac{\gamma_{\phi\phi,x}}{2\gamma_{xx}}-\frac{\gamma_{\phi\phi}^2}{(p-\gamma_{\phi})\sqrt{\gamma}}(\text{curl}\gamma)_y\nonumber
\end{eqnarray}

here 

\begin{eqnarray}
&&(\text{curl}\gamma)_x=\frac{4\alpha xy}{\sigma(\sigma-2x)}\\
&&(\text{curl}\gamma)_y=\frac{2\alpha x^2 (1-\alpha^2 x^2 y^2)}{\sigma(\sigma-2x)}\\
&&(\text{curlcurl}\gamma)_\phi=\frac{4\alpha x^4 \delta (1-y^2)}{\sigma(\sigma-2x)^2}\\
\end{eqnarray}

The equation for $y(\phi)$ can be obtained by interchanging $x$ and $y$ in all the expressions. and reverse the sign of the terms containing $\text{curl}\gamma$ or $\text{curlcurl}\gamma$.

An important issue with this framework is that there is a coordinate singularity in the equations and hence we cannot deal with the light beam being very close to the black hole or passing through the ergosphere\cite{Frolov:2024olb}. Since prograde beams (moving in the same direction as the black hole's spin) in general are close to the black hole, we focus only on the retrograde case (moving in the opposite direction to the black hole's spin).

\section{Magnification and caustic points }

The magnification is defined as the ratio of the luminosities of the light source with gravitational lensing and without gravitational lensing. It can be written as the ratio of the angular area of the light beam without or with lensing, since the photon number is conserved. The angular area of the light beam without gravitational lensing is
\begin{equation}
\Delta A_e =(D_{OL}+D_{LS})^2 \Delta\Omega_e
\end{equation}

$\Omega_e$ is the solid angle of the light beam at the source. The angular area of the light beam with gravitational lensing is 
\begin{equation}
\Delta A_o =D_{OL}^2 \Delta\Omega_o
\end{equation}
$\Delta \Omega_o$ is the solid angle of the light beam measured in the Boyer-Lindquist coordinate. The magnification is 
\begin{equation}
\label{magnification}
\mu = \frac{(D_{OL}+D_{LS})^2}{D_{LS}^2} \left|  \frac{\Delta \Omega_e}{\Delta \Omega_o} \right| 
\end{equation} 

\begin{figure}[h]
\includegraphics[width=7cm]{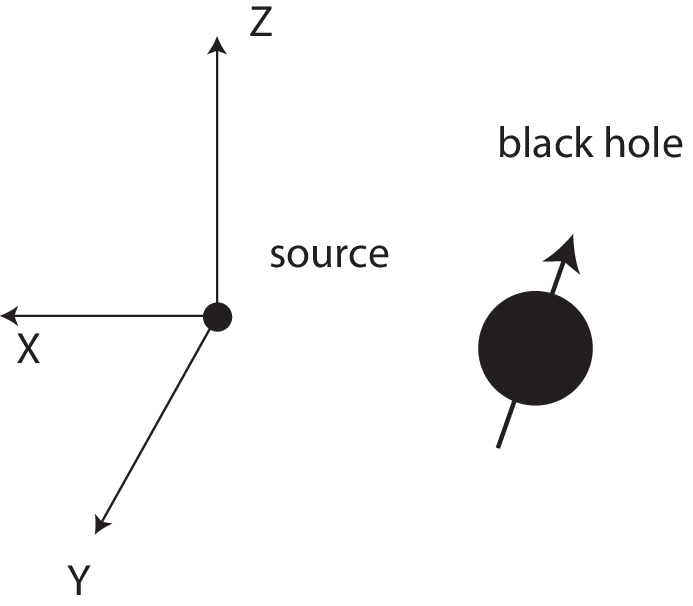}
\caption{The origin of the coordinate system is at the source. The x-direction, y-direction, and z-direction are parallel to $\hat{r}$, $\hat{\phi}$, and $\hat{\theta}$ respectively in the Boyer-Lindquist coordinate. The arrow on the black hole marks the z-direction in the Boyer-Lindquist coordinate. }
\label{coordinate}
\end{figure}

To calculate the $\Delta \Omega_o$, we must define a new coordinate in which the source is at the origin. The three axes are defined to be parallel to $\hat{r}$, $\hat{\phi}$, and $\hat{\phi}$ accordingly (Fig.\ref{coordinate}).

The velocity of the light beam in this coordinate is 
\begin{eqnarray}
V_X&=&\sin\theta_e \cos\phi_e=\frac{dr}{d l}\\
V_Y&=&\sin\theta_e \sin\phi_e=r\sin\theta \frac{d\phi}{d l}\\
V_Z&=&\cos\theta_e =r\frac{d\theta}{d l}
\end{eqnarray}

$\theta_e$ and $\phi_e$ are the corresponding spherical coordinates. We consider that both $\theta_e$ and $\phi_e$ are very small. Otherwise, the gravitational lensing effect is neglected since the light ray is far away from the black hole. The above equations are approximated to
\begin{eqnarray}
\phi_e&\approx& -x\sin\theta\frac{d\phi}{dx}\\
\theta_e-\frac{\pi}{2}&\approx& -x\frac{\frac{dy}{d\phi}}{\frac{dx}{d\phi}}
\end{eqnarray} 
The solid angle of the light beam near the source is 
\begin{equation}
\Delta \Omega_e = \sin\theta_e \Delta \theta_e \Delta \phi_e =\sin\theta_e \begin{vmatrix}
\partial_p \phi_e & \partial_p \theta_e  \\
\partial_{y_e} \phi_e & \partial_{ye} \theta_e 
\end{vmatrix} \Delta p \Delta y_e
\end{equation}
here, the Jacobian matrix is introduced to transform the variable to $p$ and $y_e$ variables. The solid angle of the light beam with respect to the black hole coordinate at $x_o$ is   

\begin{equation}
\Delta\Omega_o = \sin\theta \Delta\theta \Delta \phi =\sin\theta \begin{vmatrix}
\partial_p \phi & \partial_p \theta  \\
\partial_{y_e} \phi & \partial_{ye} \theta 
\end{vmatrix} \Delta p \Delta y_e
\end{equation}

The intensities at caustic points are enhanced infinitely. This happens while the denominator in Eq. (\ref{magnification}) is 0,
\begin{equation} 
\Delta \Omega_o=0 
\end{equation}

\begin{figure}[h]
\includegraphics[width=11cm]{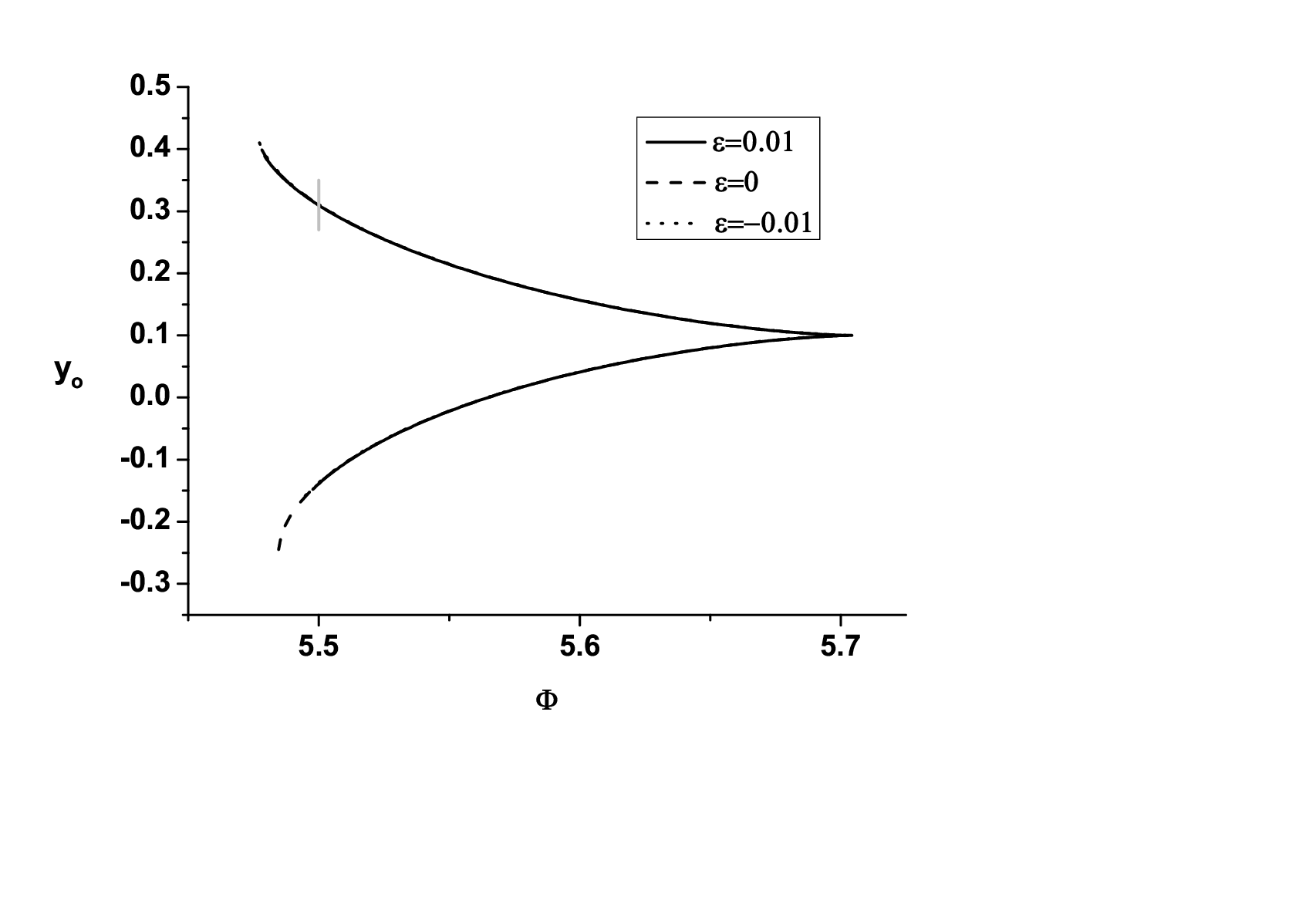}
\caption{ The secondary caustics on the celestial sphere: The light source is at radius $r=10^{10} M$, $\Phi_e=0$ and $y_e=0$. The spin parameter of the black hole is $a=-1$. The observer is at $r=10^9 M$. $\epsilon$'s are chosen accordingly. The gray vertical line marks the enlarged region shown in Fig.\ref{y=0.1-enlarge}.}
\label{y=0.1}
\end{figure}

\begin{figure}[h]
\includegraphics[width=11cm]{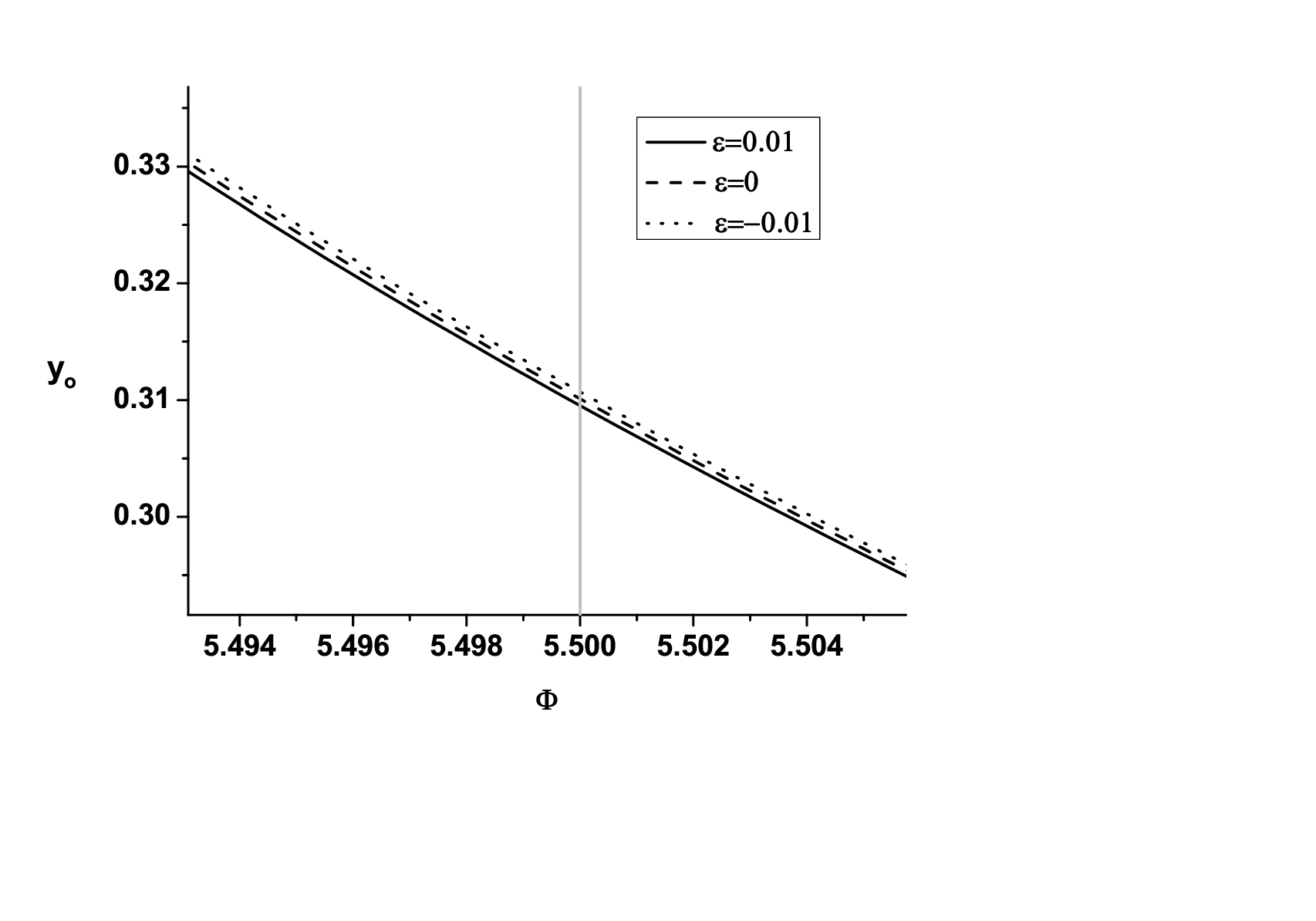}
\caption{The secondary caustics on the celestial sphere: The light source is at radius $r=10^{10} M$, $\Phi_e=0$, and $y_e=0.1$. The spin parameter of the black hole is $a=-1$. The observer is at $r=10^9 M$. $\epsilon$'s are chosen accordingly. The gray vertical line marks $\Phi_o=5.5$ circle.}
\label{y=0.1-enlarge}
\end{figure}

\begin{figure}[h]
\includegraphics[width=11cm]{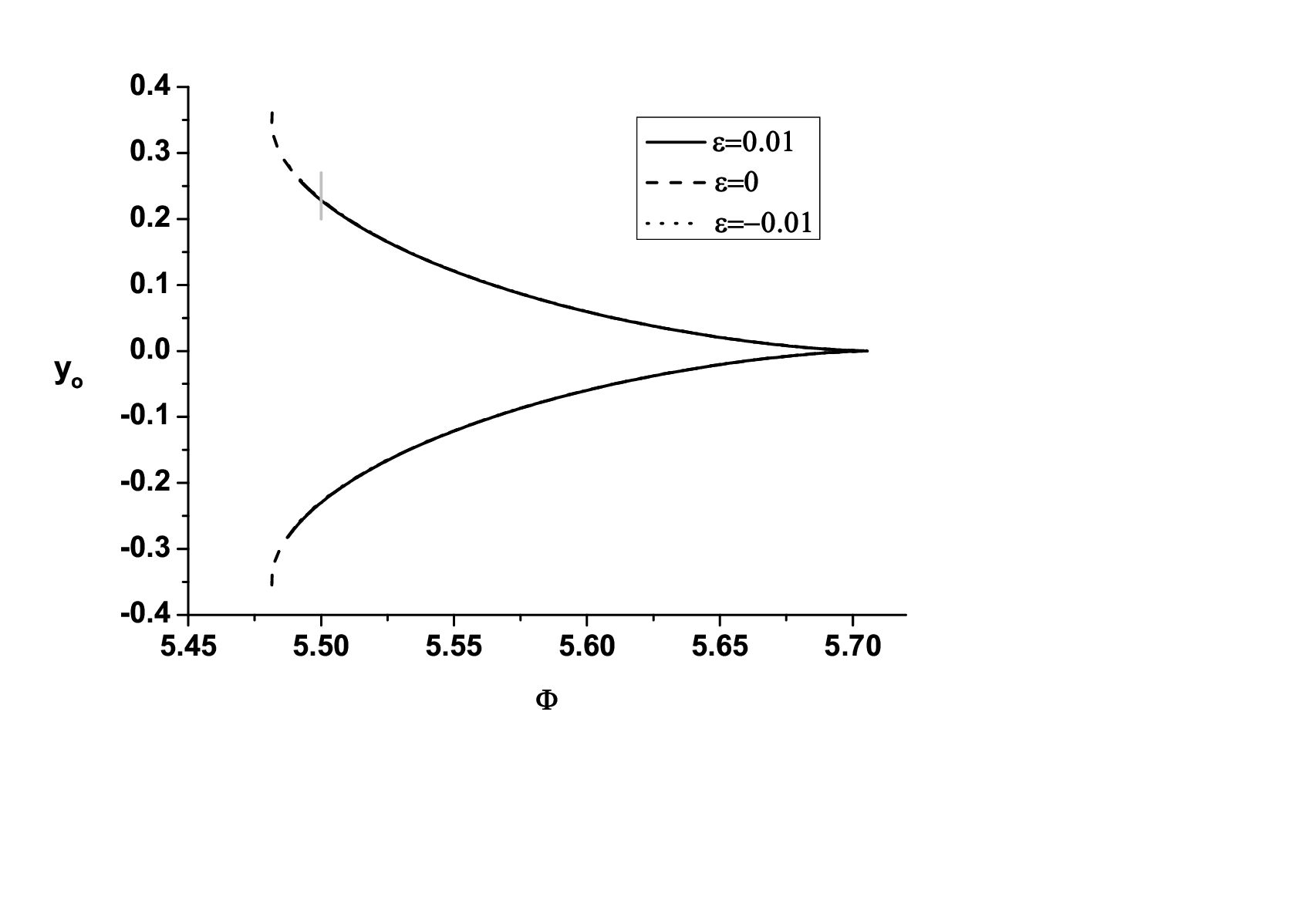}
\caption{The secondary caustics on the celestial sphere: The light source is at radius $r=10^{10} M$, $\Phi_e=0$ and $y_e=0$. The spin parameter of the black hole is $a=-1$. The observer is at $r=10^9 M$. $\epsilon$'s are chosen accordingly. The gray vertical line marks the enlarged region shown in Fig.\ref{y=0-enlarge}. }
\label{y=0}
\end{figure}

\begin{figure}[h]
\includegraphics[width=11cm]{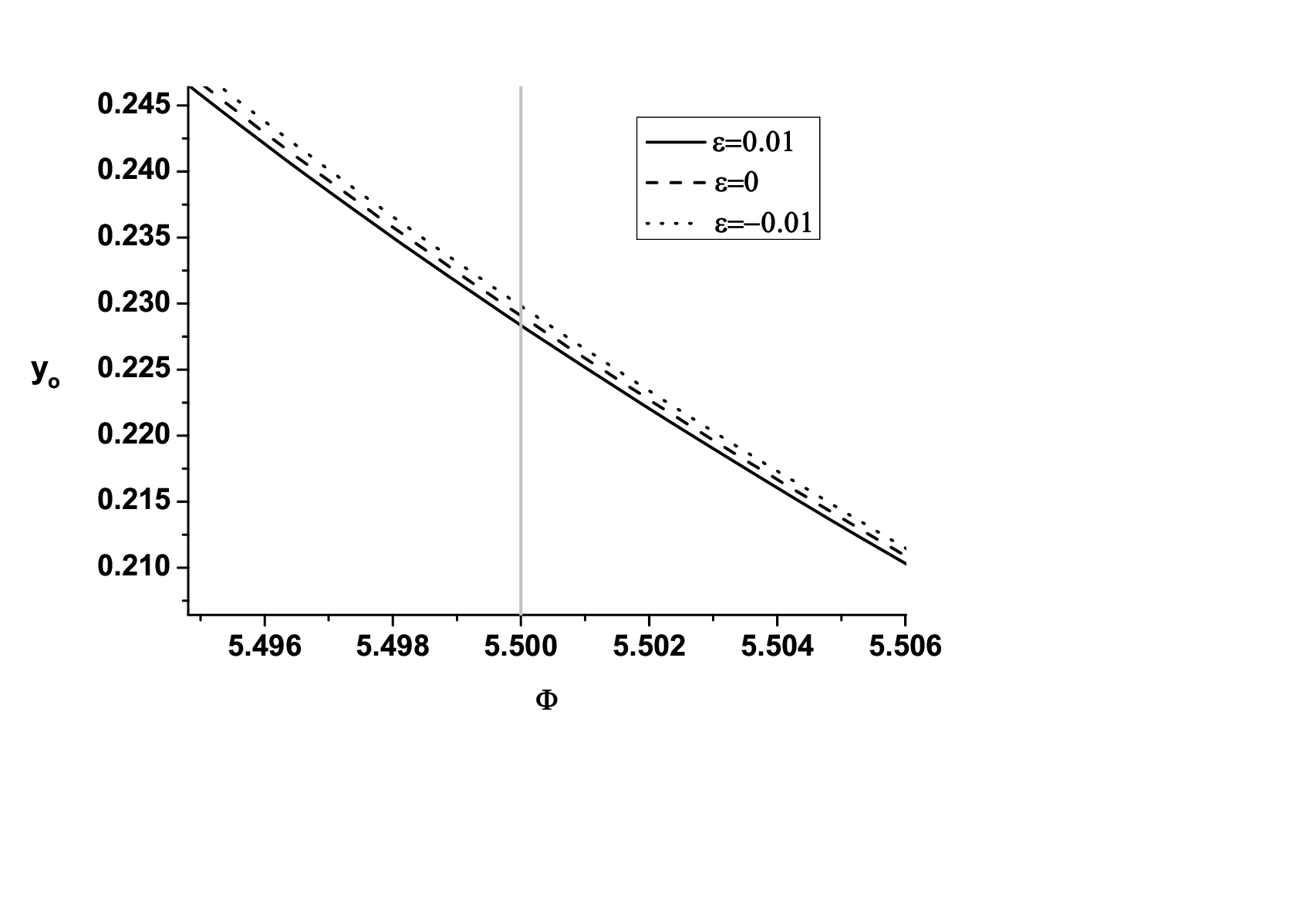}
\caption{The secondary caustics on the celestial sphere: The light source is at radius $r=10^{10} M$, $\Phi_e=0$, and $y_e=0$. The spin parameter of the black hole is $a=-1$. The observer is at $r=10^9 M$. $\epsilon$'s are chosen accordingly. The gray vertical line marks $\Phi_o=5.5$ circle.}
\label{y=0-enlarge}
\end{figure}

To enhance the black hole's spin effect, we choose the black hole spin parameter to be $a=-1$. For negative $a$, the retrograde beam is circulating in positive $\hat{\phi}$ direction. Fig.\ref{y=0.1} and \ref{y=0} show the second caustic points of the retrograde light for the observers. The observer at this point sees an infinite enhancement of the brightness of the point source. The sources of Fig.\ref{y=0.1} and Fig.\ref{y=0} and are at $(r=10^{10}M,y=0.1,\phi=0 )$ and $(r=10^{10}M,y=0,\phi=0 )$ respectively. The observer is at $r=10^9M$. One can see that if the source is at the north sphere, the caustics are also deviated to the north. This is the same as previous studies\cite{Bozza:2008mi,Bozza:2006nm,Bozza:2005kq,Bozza:2002af,Gyulchev:2006zg,Bozza:2004kq,Sereno:2007gd}. In these figures, the spin effect of electromagnetic fields is small so that the curves overlap. The enlarged versions are shown in Fig. \ref{y=0.1-enlarge} and \ref{y=0-enlarge}. The $\epsilon=0.01$ is moving down and $\epsilon=-0.01$ is moving up compared with the no spin case $\epsilon=0$.

\section{Possible observation}

\begin{figure}[h]
\includegraphics[width=11cm]{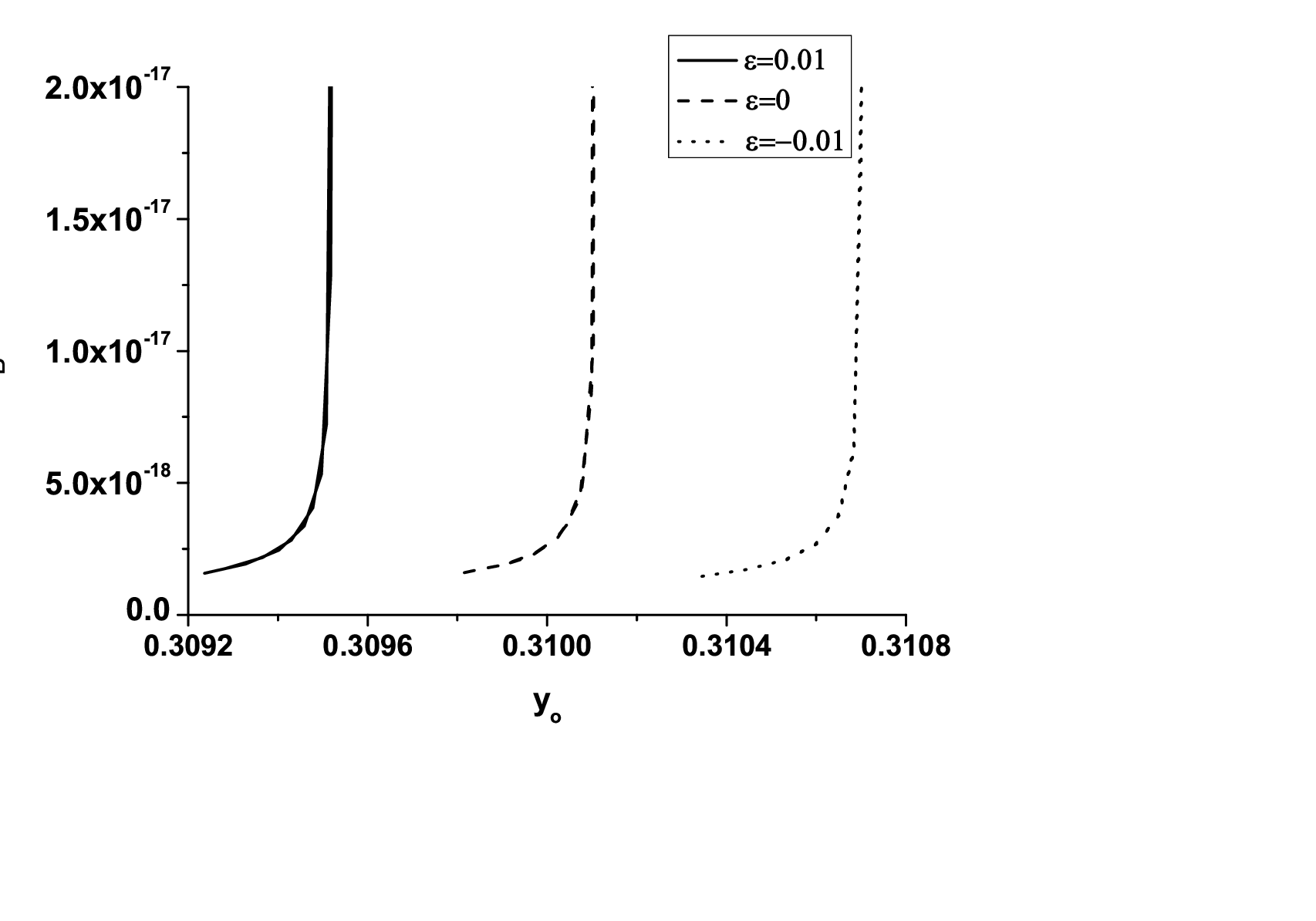}
\caption{The enhancement curve: The light source is at radius $r=10^{10} M$, $\Phi_e=0$ and $y_e=0.1$. The spin parameter of the black hole is $a=-1$. The observer is at $r=10^9 M$. $\epsilon$'s are chosen accordingly. The observer is at $\Phi_o=5.5$ and is marked as a gray line in Fig.\ref{y=0.1}. The factor $\frac{(D_{OL}+D_{LS})^2}{D_{LS}^2}$ is neglected in the plot.}
\label{factor-y=0.1}
\end{figure}

\begin{figure}[h]
\includegraphics[width=11cm]{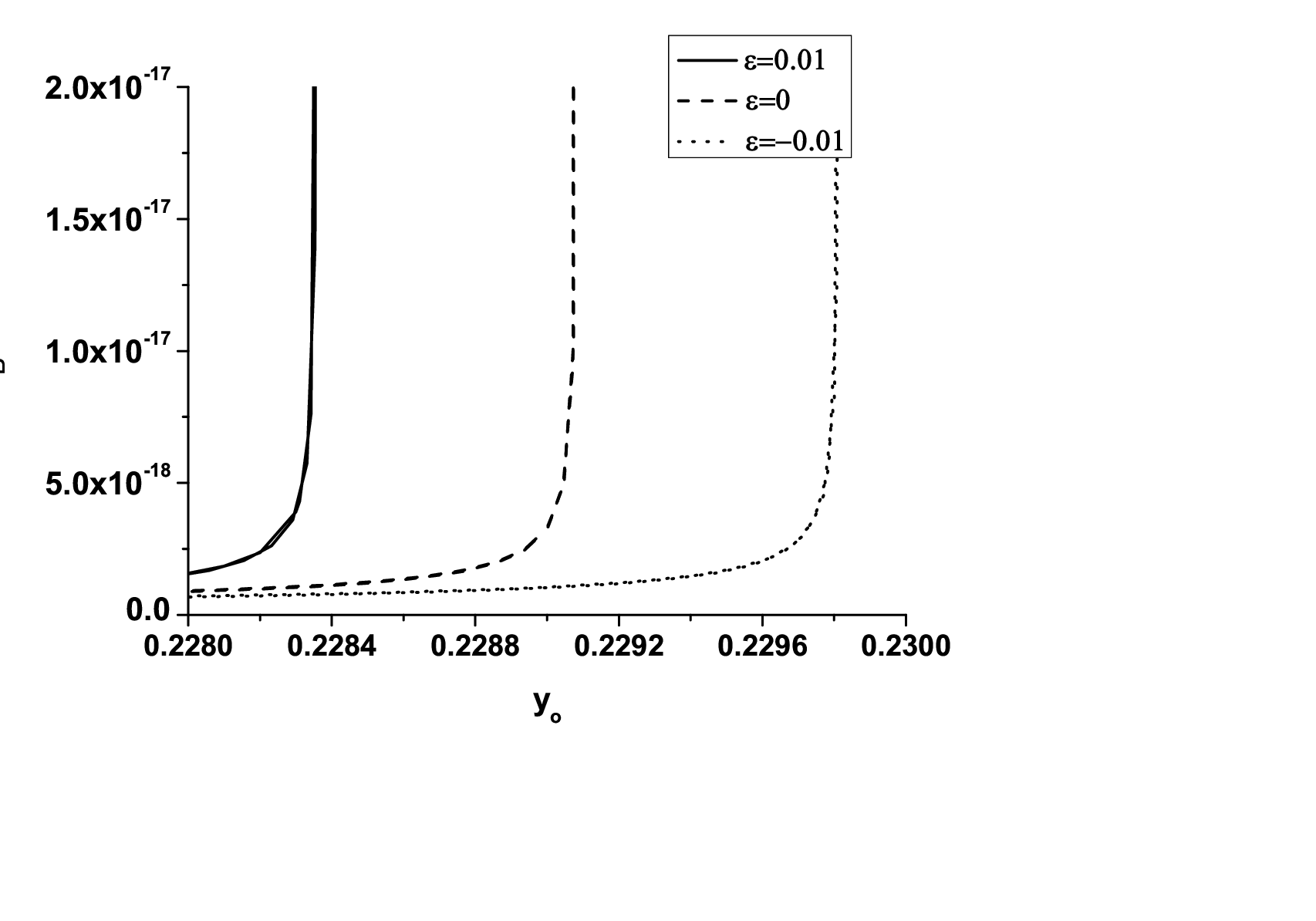}
\caption{ The enhancement curve: The light source is at radius $r=10^{10} M$, $\Phi_e=0$ and $y_e=0$. The spin parameter of the black hole is $a=-1$. The observer is at $r=10^9 M$. $\epsilon$'s are chosen accordingly. The observer is at $\Phi_o=5.5$ and is marked as a gray line in Fig.\ref{y=0}. The factor $\frac{(D_{OL}+D_{LS})^2}{D_{LS}^2}$ is neglected in the plot}
\label{factor-y=0}
\end{figure}

It has been noticed that a Kerr black hole can split the images  due to their own polarization, but the effect depends on the sensitivity of the devices\cite{Frolov:2024olb}. Here we have shown that the caustics are also split due to the polarization. Caustics are the points where infinite magnification of the luminosity occurs. Therefore, an observer at such a location overcomes the sensitivity issue. Fig.\ref{factor-y=0.1} and \ref{factor-y=0} show the relative magnification, while an observer moves along the $\Phi=5.5$rad. One can see that both $\epsilon=0.01 $ and $\epsilon=-0.01 $ are deviated from non-polarization light, $\epsilon=0 $. Only magnifications around caustics are plotted in these figures. Therefore, one cannot see the magnifications for $y_e$ larger than the location of the infinite peaks.  It is also important that every single curve is the overlap of two very close curves. Therefore, the enhancement is about double the amount shown in the figures. Fig.\ref{factor-y=0.1} and \ref{factor-y=0} show that the splitting for two different polarized beams is about $\Delta y_e\approx 10^{-3}$. Since the radius of the observer and the source are far larger than the black hole radius, the splitting will not change much for even farther observers or sources.
 
Consider an isolated black hole of mass about $M_\odot$ that appears at several light years away from the Earth, then the splitting is about $10^{12}$m. While the Earth orbits the sun, the detector will see two distinct peaks due to the light polarizations(one for left circular polarization and one for right circular polarization). This phenomenon is distinguishable from the non-polarized  case, $\epsilon=0$, and can be a hint to the spin-optics effect. Since the light is polarized at the caustics, one may search for a short-period circularly polarized light. One must be careful that our calculation is based on a point source. If a source's angular diameter is much larger than $10^{-4}$, the brightness change must be averaged according to the source's radius.

Since $\epsilon=\pm \frac{1}{2\omega M}$, the calculation for splitting polarized light is correct only for light wavelengths much smaller than the black hole radius. In this study we choose $|\epsilon| = 0.01$. For a solar mass black hole, whose radius is about $10$km, the target wavelength of light must be about or less than $0.1$km. This signal will be a suddenly appearing and disappearing polarized light. This suddenly appearing signal is similar to WOW signal\cite{Kraus}. Even though the WOW signal is definitely not from a black hole scattering, the study strategy is similar. The difference is that the brightness becomes strong twice and these two peaks have different polarizations. 
A visible light source can be a tool to study much smaller black holes whose mass and radii are about $10^{-8}M_\odot$ and 100$\mu$m respectively. If a such massive primordial black hole passes Earth from $1000R_\oplus $ away, two strong peaks are separated about $R_\oplus$. They are well separated and distinguishable. Therefore, the problem is whether we have enough time and a strong background light source to identify the polarized light. A possible source coud be a closeby supernova.

\section{conclusion}

Gravitational lensing in strong gravitational fields is an appealing phenomenon within general relativity. Potentially, it can be a probe for physics in a region very close to a black hole horizon, where the effect of the spin of a black hole becomes important. When light passes very close to the horizon, it can orbit the black hole several times before it is scattered away. The intensity of the image is generally weak. However, if the observer happens to be near the caustic points, the light is convergent and the intensity is highly enhanced. This gives a possible method to study the strong gravitational effect by gravitational lensing. Both microlensing and Sgr A* can be potential lensing exmaples\cite{Holz:2002uf,DePaolisi2003,Eiroa:2003jf,DePaolis:2004xe}.

Recently, more and more studies have shown that spinning particles propagate along a path deviating from the geodesic. An unpolarized light beam with right and left circular polarization will be split into two circular polarized beams after passing a strong gravitational region\cite{Frolov:2012zn}, as Fig.\ref{light-path} shows. Whether this splitting can be detected depends on the splitting angle and the intensities of these beams. To overcome the difficulty, we propose studying the retrolensing near the caustics. The intensity of a point light source is infinitely enhanced at caustics, which are separated according to the polarization of light. An extremal black hole splits the caustics point up to $10^{-3}$ rad as shown in Fig.\ref{y=0.1-enlarge} and \ref{y=0-enlarge}. This small angle becomes a large distance after propagating a long distance and becomes distinguishable. An observer may see different polarized beams appearing at different locations sequentially as shown in Fig.\ref{factor-y=0.1} and \ref{factor-y=0}.  This is a very strong hint that an unpolarized light is scattered by a Kerr black hole.

\begin{acknowledgments}
D.C. Dai is supported by the National Science and Technology Council (under grant no. 114-2112-M-259-008-MY3).    
\end{acknowledgments}

\end{document}